\documentclass[dvips]{article}
\usepackage{icrctc07}

\title{\boldmath Gamma-Ray, Neutrino \& Gravitational Wave Detection: \\ OG 2.5,2.6,2.7 Rapporteur}
\shorttitle{Rapporteur OG 2.5,2.6,2.7}
\authors{G.~Rowell}
\shortauthors{G. Rowell}
\afiliations{School of Chemistry \& Physics, University of Adelaide, Adelaide 5005, Australia}

\abstract{
This report is based on a rapporteur talk presented at the 30th International Cosmic Ray Conference
held in Merida, Mexico (July 2007), and covers three of the OG sessions devoted to neutrino,
gravitational wave, and $\gamma$-ray detection.}

\email{growell@physics.adelaide.edu.au}

\begin{document}
\maketitle

\section{\boldmath Introduction \& Overview}

Summarised here are key results from papers and posters presented in the sessions OG~2.5 (neutrino detection), 
2.6 (gravitational wave detection), and 2.7 ($\gamma$-ray detection). 
The number of presentations in each session was
75 (OG~2.7), 19 (OG~2.5) and 2 (OG~2.6), with $\gamma$-ray detection clearly dominating. 
OG~2.5 was devoted to neutrino/$\gamma$-ray
connections and related experimental and theoretical issues and contains overlaps with the HE~2 session. 
A more detailed summary of neutrino detectors and related astrophysical theory can be found
in the HE~2 rapporteur by Tom Gaisser \cite{Gaisser-RAPP}. Here, summaries of each sessions are 
ordered according to the number of contributions. 
I have kept to citing only ICRC presentations/posters since in most cases
a detailed list of references may be found therein.

\section{\boldmath OG~2.7: Gamma-Ray Detection}
This $\gamma$-ray detection session encompassed technical status reports from space and ground $\gamma$-ray instruments, 
summaries of their analysis techniques and performance, and plans for future instruments. Results from $\gamma$-ray instruments
(sessions OG~2.1 to 2.4) are summarised by Jim Hinton \cite{Hinton-RAPP} and I will only touch on a few key relevant results.

\subsection{Ground-Based $\gamma$-Ray Detectors: Current \& Funded}

Ground-based $\gamma$-ray detectors can be broadly split into three camps:  
(i) Telescopes employing the Imaging Atmospheric Cherenkov Technique (IACT) including those with $\geq$2 telescopes operating
as stereoscopic arrays, (ii) Water Cherenkov detectors, and (iii) Ground arrays.

\subsubsection{IACT Detectors/Arrays}
There are four major IACT detectors/arrays in operation today, and status reports 
presented at this conference on the two most recently commissioned systems --- VERITAS and MAGIC/MAGIC-II, were dominant in
number. Both H.E.S.S. and CANGAROO-III have been in full operation
since $\sim$2004 and much of their technical details have been presented at previous ICRCs, although an updated status of CANGAROO-III
was presented here.

\begin{figure}[t]
  \centering
  \includegraphics[width=0.45\textwidth]{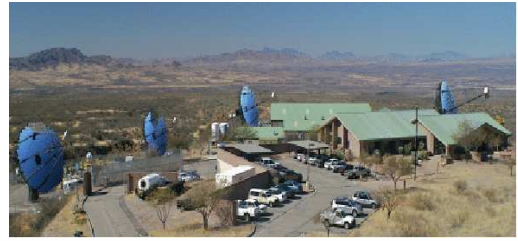}
  \includegraphics[width=0.5\textwidth]{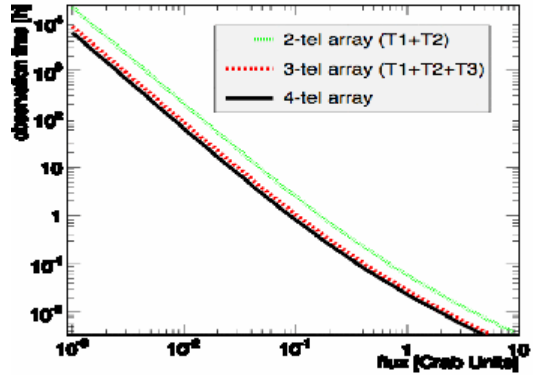}
  \caption{{\bf Top:} View of the VERITAS 4-telescope IACT array \cite{Maier:1}; {\bf Bottom:} VERITAS 
    flux sensitivity (Crab units) of the VERITAS array for several telescope (T1, T2, etc..) combinations \cite{Maier:2}.}
  \label{fig:veritas}       
\end{figure}

After many (non-scientific) delays, VERITAS \cite{VERITAS} achieved first light with four telescopes in April 2007.  
\cite{Maier:1} presented an overview of the VERITAS array
which comprises 4$\times 106$m$^2$ telescopes situated at the Basecamp at Fred Lawrence Whipple Observatory, Mt. Hopkins, Arizona 
(Fig~\ref{fig:veritas}). The optics of each telescope comprises a Davies-Cotton dish with 345 mirror segments \cite{Roache:1}.
Mirror mis-alignment due to dish deformation vs. elevation has been successfully corrected using a laser alignment 
system \cite{Toner:1}. The telescopes' cameras comprise a 499 photomultiplier (Phillips XP 2970/02 PMT) pixel array \cite{Nagai:1}, 
providing a $\sim 3.5^\circ$ field of view (FoV). The size (diameter) of each pixel is 0.15$^\circ$. The trigger system is based
on three levels: L1 - pixel; L2 - camera pixel pattern; L3 - array trigger \cite{Weinstein:1}. A pixel trigger of 4--5~pe. 
(photoelectrons) is first applied. The camera is triggered when (presently) 3 adjacent pixels within a pre-defined group
are triggered within 6~ns (see \cite{White:1} for details). This pattern logic greatly reduces accidental triggers due to skynoise.
The inter-telescope array trigger (L3) is met whenever 
$\geq$2 telescopes trigger within a time window up to 125~ns, depending on the zenith and azimuth angle of observations. 
The 3-telescope array trigger rate is $\sim$220~Hz with a 10\% dead time. PMT pulses are digitised
using 500~MHz (VME-based) flash ADCs (FADCs) \cite{Hays:1}. Several methods based on FADC sampling and filtering 
to extract the Cherenkov pulse arrival time were evaluated by \cite{Cogan:1}, with a combination of resampling and 
linear interpolation yielding a $\sim 0.2$~ns time resolution. Calibration issues encompassing the use of single-pe runs, muon ring data
and flat fielding were summarised by \cite{Hanna:1}. \cite{Hui:1} also discussed the use of laser shots as a way to 
correct for local atmospherics.
A data analysis chain outlined by \cite{Daniel:1} focused on the specific eventdisplay and VEGAS \cite{Cogan:2} packages. 
Together they handle the image formation (from FADCs), image calibration, cleaning (employing a well-known picture/boundary philosophy) 
and stereo reconstruction of event direction. 
As for any new instrument such as this, observations of the Crab provide the first real test and results have clearly met expectations 
gleaned from earlier Monte-Carlo (MC) studies \cite{Maier:2} (see Fig.~\ref{fig:veritas}). Overall the VERITAS array provides an 
eventwise angular 
resolution better than 0.14$^\circ$ and a 5$\sigma$ detection of a 10\% Crab flux in under 1 hour. 
With all of these results, it's now clear that VERITAS is fully functioning and we eagerly await the new high energy astrophysics
to come (see \cite{Hinton-RAPP} for a summary of first VERITAS results). 

\begin{figure}
  \centering
  \includegraphics[width=0.48\textwidth]{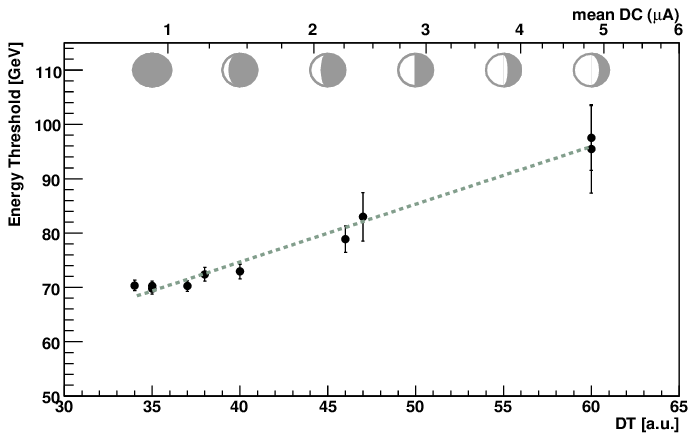}
  \includegraphics[width=0.45\textwidth]{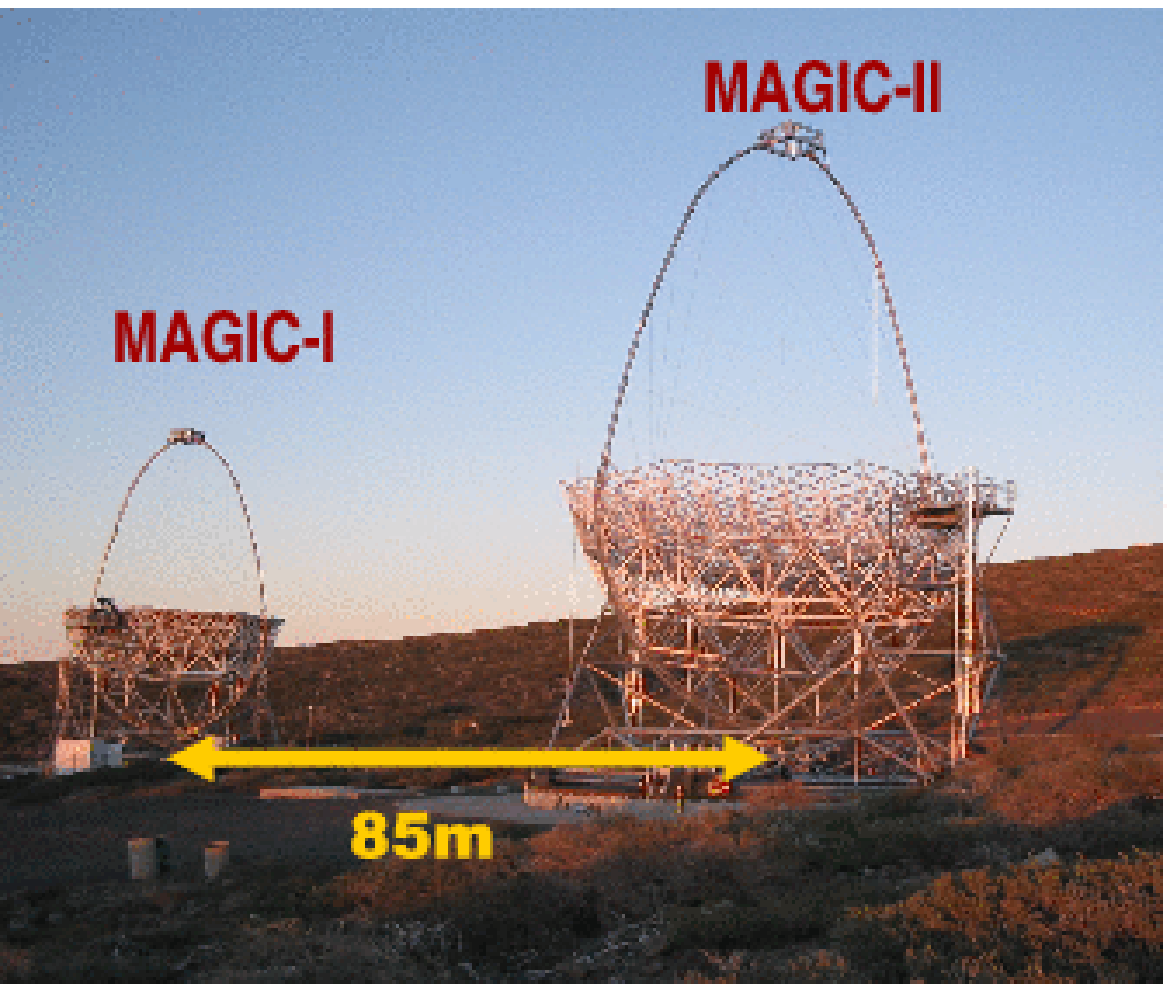}
  \includegraphics[width=0.53\textwidth]{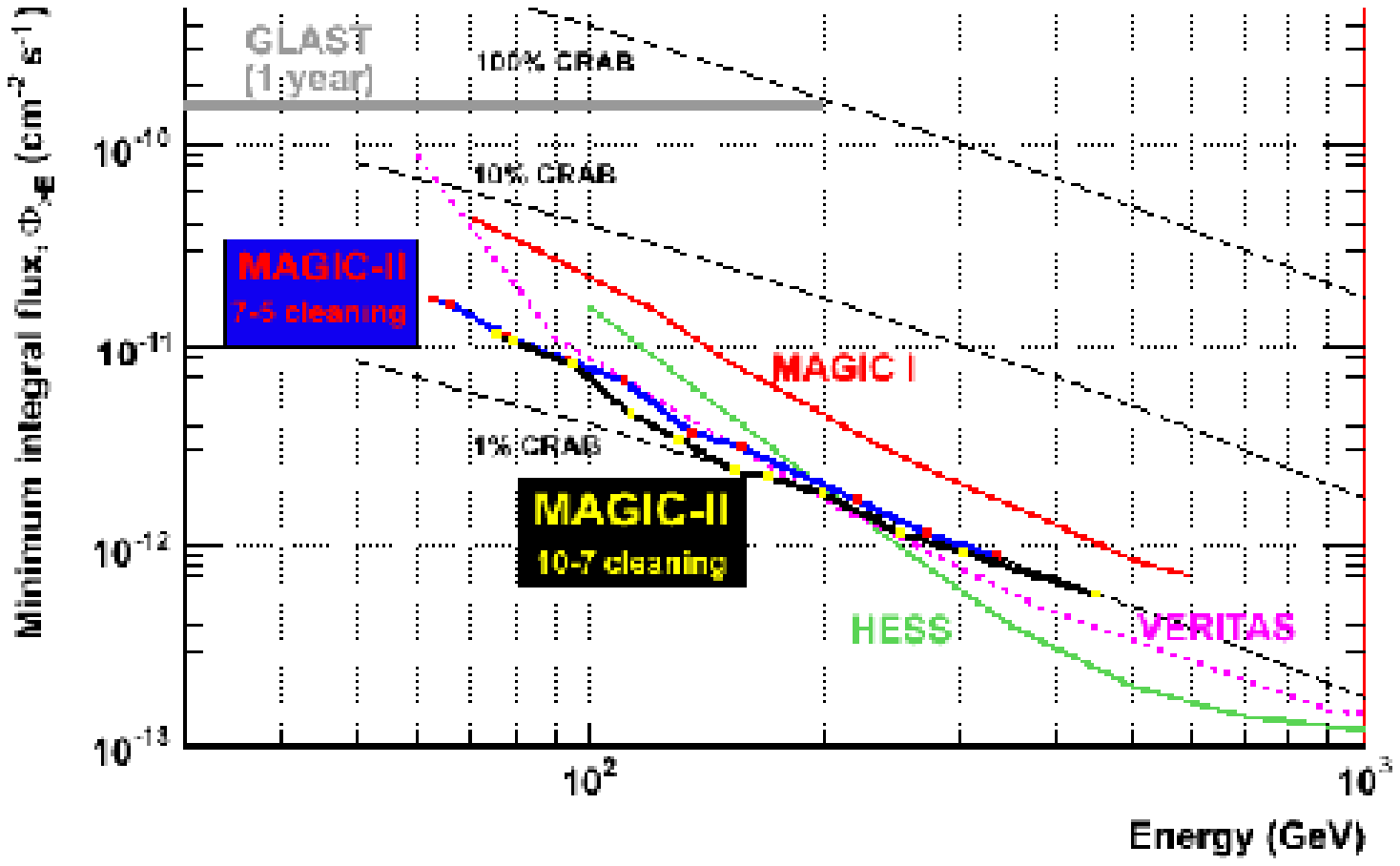}
  \caption{{\bf Top:} Variation of MAGIC energy threshold vs. pixel discriminator threshold (DT) and Moon phase \cite{Rico:1}.
    {\bf Middle:} Picture of the two MAGIC telescopes -- MAGIC-II \cite{Goebel:1}. {\bf Bottom} MAGIC-II flux sensitivity
    (for two versions of image cleaning) compared with that from other $\gamma$-ray telescopes \cite{Goebel:1}.}
  \label{fig:magicII}       
\end{figure}

The MAGIC telescope \cite{MAGIC} has been in full operation since 2004 and results have flowed steadily since. 
Technical issues presented at this conference were devoted to updates of the MAGIC camera \& electronics, improvements
to data analysis, and the status of phase II (MAGIC-II) of the project in which a second telescope will soon be operational.
An overview of MAGIC/MAGIC-II was presented by \cite{Goebel:1}. MAGIC is a single large (236~m$^2$) telescope
situated at the 2200~m a.s.l  on the Canary Island of La~Palma. The MAGIC team have emphasised the use of new technology 
such as lightweight construction (for fast slewing, $\sim$40~s, motivated by GRB followups), optical fibre data transmission, 
and research into high quantum efficiency PMTs and other photon detectors. The latter is motivated by the push to reduce the 
energy threshold of ground-based $\gamma$-ray instruments to well below 100~GeV. Presently the trigger threshold is 
of MAGIC $\sim 50$~GeV and the 5$\sigma$ flux sensitivity for 50~h observation is $\sim$2\% Crab for energies above 100~GeV.
Effort has gone into development of hadronic background rejection and one method summarised by \cite{Ferenc:1}
based on utilising the Cherenkov light present in hadronic components of cosmic-ray (CR) showers, which leads to an increased number
of pixels with small (1--1.5 pe.) signals compared to $\gamma$-ray images, 
indicated improved hadron rejection by a factor $\sim2.6$ (Crab excess significance $\sim$9 to 12$\sigma$) for energies $E<200$~GeV.
The lightweight nature of the MAGIC dish has meant that active mirror control is necessary and \cite{Biland:1} described the
successful method of mirror re-alignment. Using a pre-determined database of mirror positions vs. azimuth and altitude (derived
from laser and star measurements), the 
re-alignment procedure can be completed in $\sim$10~s, in parallel with telescope slewing such that the re-alignment adds no
additional time delay. Extending the duty cycle of ground-based $\gamma$-ray instruments has also been a focus with
observations running into bright Moon phases and even into twilight have been investigated \cite{Rico:1}. 
With this in mind the MAGIC camera PMTs are operated at lower gains, few$\times 10^4$ as opposed to the $\sim 10^6$ values 
traditionally used.
In order to control accidental triggers the pixel discriminator threshold (DT) was increased with Moon brightness, with a
resultant increase in energy threshold (Fig.~\ref{fig:magicII}). The corresponding reduction in event statistics extends up 
to image {\em size}$\sim$10$^4$, however no strong changes in distributions of image {\em length} and {\em width} for 
{\em size}$\geq$400~pe. was noticed suggesting that the hadron rejection cuts can remain essentially unchanged at these higher
thresholds. Improvements (Feb. 2007) to the MAGIC electronics with multiplexed 2~GHz FADCs (replacing the previous 300~MHz system) 
have also indicated improved hadron rejection based on timing differences between $\gamma$-ray and hadron showers 
\cite{Goebel:2,Tescaro:1}.
MAGIC-II will see an additional telescope constructed 85~m away from the first MAGIC telescope. The new dish and mount are in place
(Fig.~\ref{fig:magicII}) and first light is expected in the first half of 2008. Several improvements are foreseen in the second telescope.
FADC type electronics based on low power switched capacitor ring buffers (Domino Ring Buffer - not unlike the Analogue Ring 
Samplers in use by H.E.S.S.) are planned, and have been successfully tested on site. The camera, comprising 1039$\times$0.1$^\circ$ 
pixels will subtend a FoV 3.5$^\circ$, similar to the MAGIC-I camera, but have a larger trigger area \cite{Hsu:1}. After initially 
equipping the camera with QE$\sim$30\% PMTs, higher QE ($\sim$50\%) PMTs are planned \cite{Hsu:2}. Mirror segments about 4 times
larger in area (1~m$^2$ area) compared to MAGIC-I are also installed \cite{Bastieri:1}.
Overall, with the advantage of stereoscopy, the MAGIC-II system is expected to operate with sensitivity a factor $\sim$3 better
and energy threshold $\sim$40\% lower than the single telescope (Fig.~\ref{fig:magicII}). 
Looking further afield, research into avalanche photodiodes
(APDs) continues and \cite{Otte:1} summarised field trials of prototype Hamamatsu APD arrays operated in Geiger mode. 
APDs can achieve QE of 60\% or more and present a way to considerably reduce the energy threshold
of $\gamma$-ray telescopes, despite the drawbacks of small (mm) size, high cross talk and dark current. After
initial tests on a solar concentrator, they were installed on the MAGIC camera for several nights. 
Cherenkov signals from the installed APDs were obtained and a signal ratio (compared to adjacent conventional
PMTs of the MAGIC camera) of about 1.6 was indicated, consistent with the improvement in QE. Additionally, single and multiple
pe.peaks were resolved using low output laser runs. 

The status of the 4-telescope CANGAROO-III system \cite{CANGAROO} was summarised by \cite{Mori:1} with a focus on astronomical results.
Presently only three of the four telescopes operate with the same camera type and are used in stereo operations. Data
analysis is based on the Fisher discriminant derived from a linear combination of image parameters such as 
{\em width} and {\em length}. Funding for upgrades to the telescope mirrors (to improve the overall angular resolution) 
and T1 electronics (to bring T1 into the stereo trigger) have been sought.

\begin{figure}
  \centering
  \includegraphics[width=0.45\textwidth]{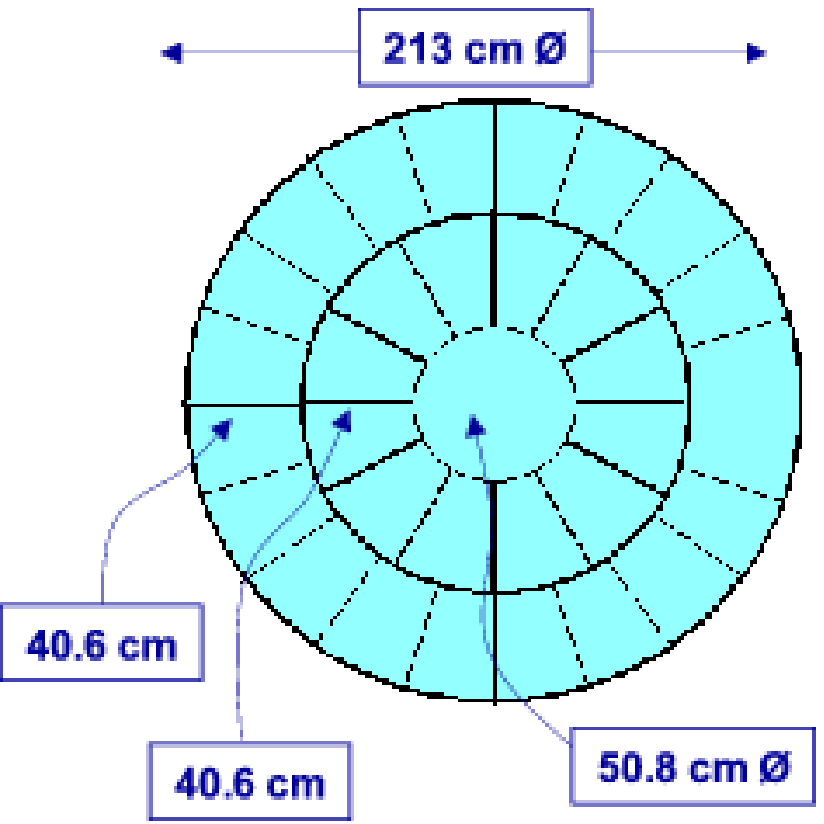}
  \includegraphics[width=0.4\textwidth]{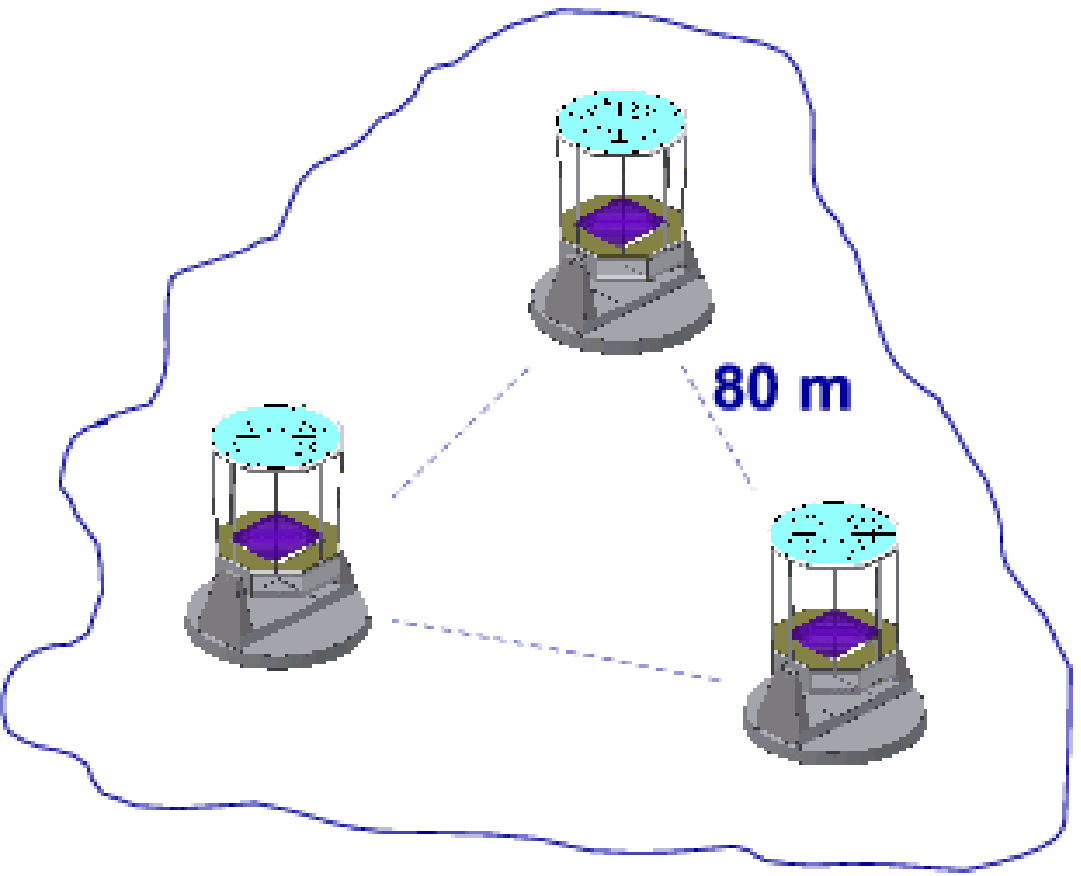}
  \includegraphics[width=0.45\textwidth]{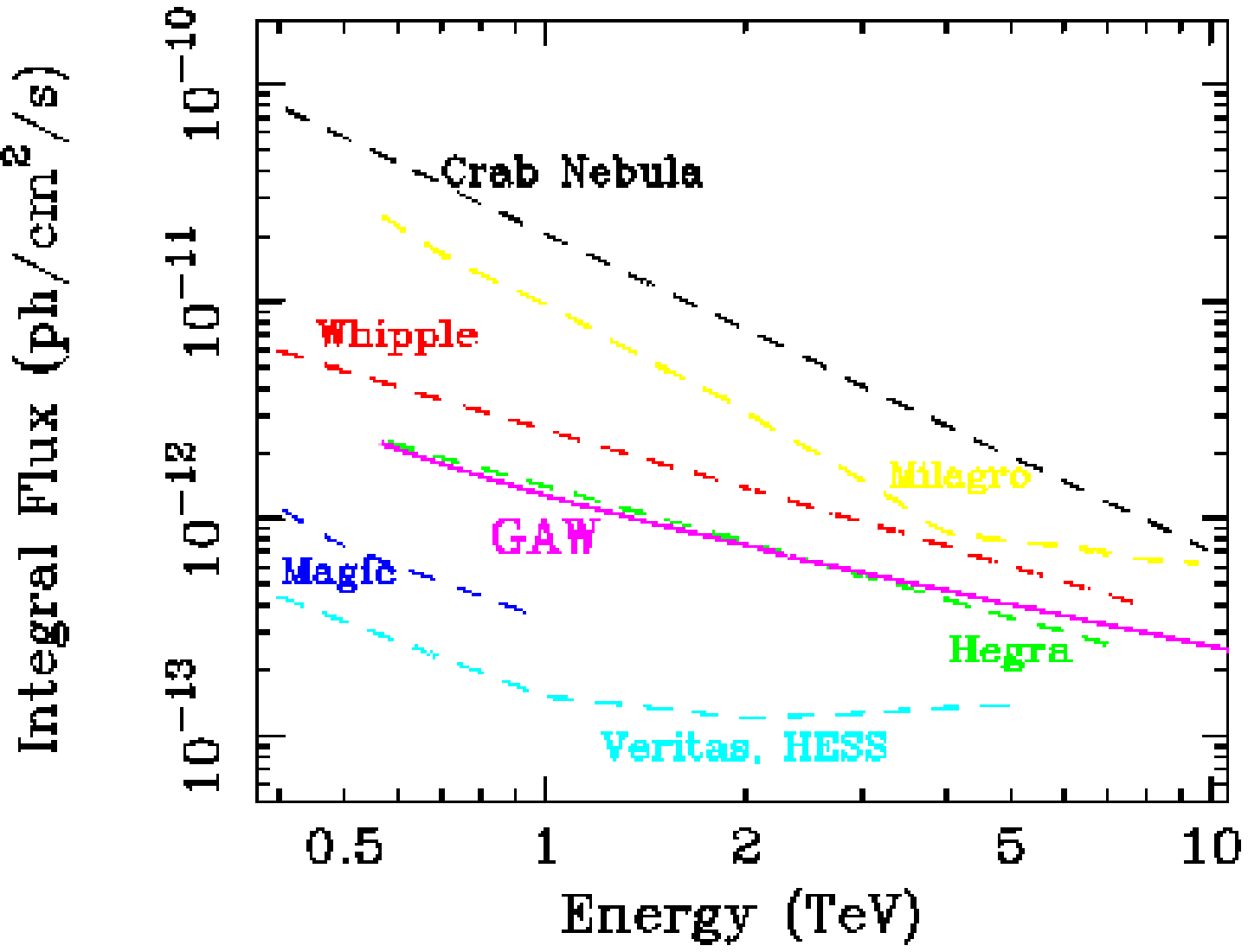}
  \caption{{\bf Top:} Fresnel lens design for GAW.
    {\bf Middle:} Layout of the GAW 3-telescope array {\bf Bottom} GAW phase-II flux sensitivity (5$\sigma$ in 50~h). 
    Plots are taken from \cite{Cusumano:1}.}
  \label{fig:gaw}       
\end{figure}

The GAW (Gamma AirWatch) project summarised by \cite{Cusumano:1,Maccarone:1} is the first serious attempt to operate an IACT
system with refractive optics, in this case a Fresnel lens. A key point is that the Fresnel lens permits a very wide FoV
up to 24$^\circ$ in diameter to be employed for wide field surveys. The GAW telescopes consist of tiled multi-anode PMTs (MAPMT)
on alt/az mounts coupled to the Fresnel (2.13~m diameter f/1.2) optic system. The Fresnel lens (Fig.~\ref{fig:gaw}) is
of a tessellated design from Fresnel Technologies Inc. (Ft. Worth, Texas). The Hamamatsu R7600-03-M64 (64 pixel) MAPMTs have 
pixel sizes of $\sim$4~arcmin. Phase-I (under construction at Calar Alto 2150~m a.s.l.) will see three identical telescopes 
arranged in a triangle of side 80~m (Fig.~\ref{fig:gaw}), each equipped with 5$^\circ \times 5^\circ$ FoV cameras. Phase-II 
will see the cameras expanded to cover  24$^\circ \times 24^\circ$. The small pixel sizes mean that they are essentially 
photon limited and thus can operate without expensive ADCs/FADCs. Simulations so far suggest an energy threshold of about 
0.5~TeV and angular resolution 0.3$^\circ$ to 0.1$^\circ$, improving with energy. The expected point source 
flux sensitivity of GAW (Phase~II) is also presented in Fig.~\ref{fig:gaw} and appears to match that of the HEGRA IACT-System.

\begin{figure}
  \centering
  \includegraphics[width=0.45\textwidth]{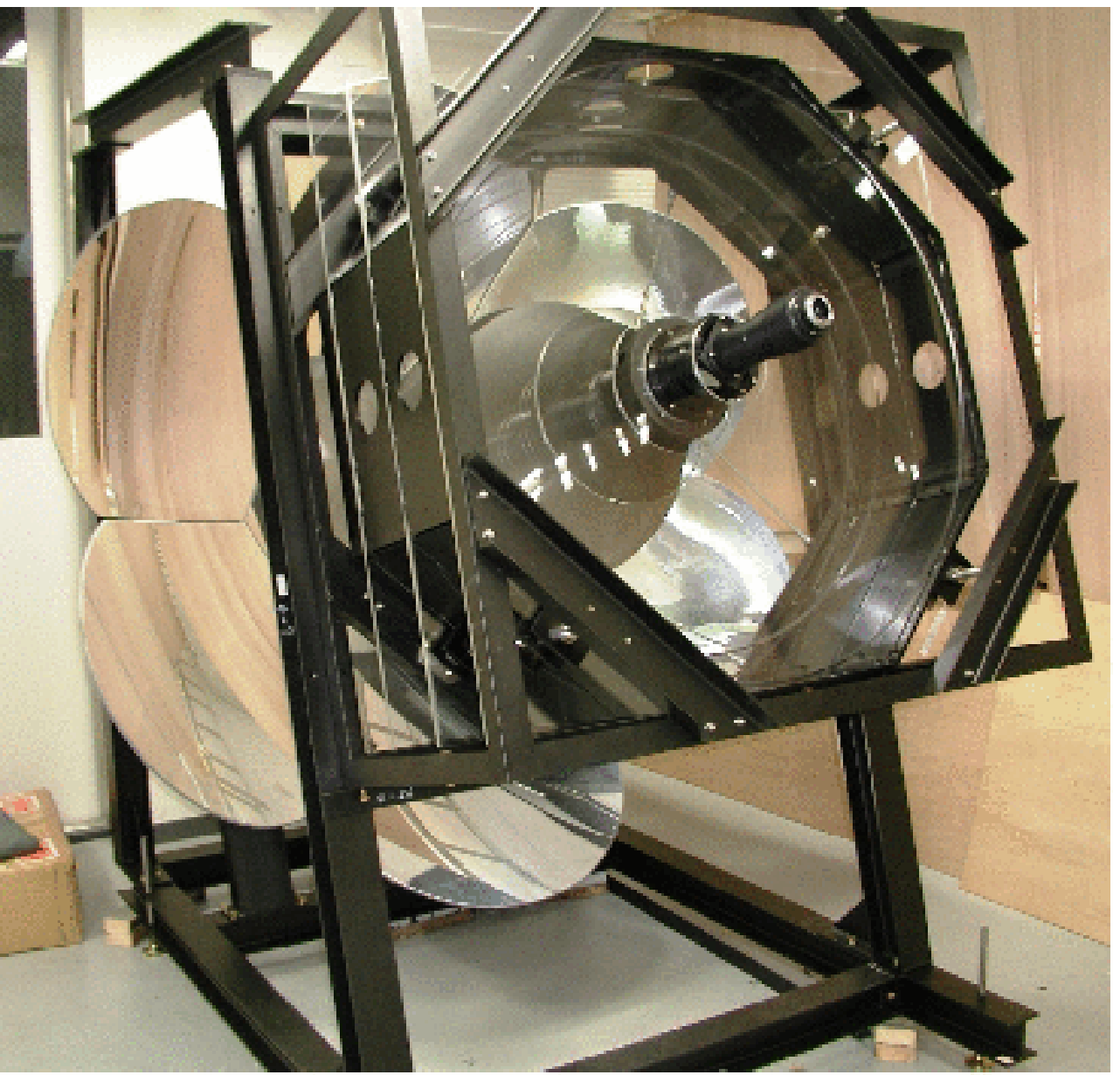}
  \includegraphics[width=0.45\textwidth]{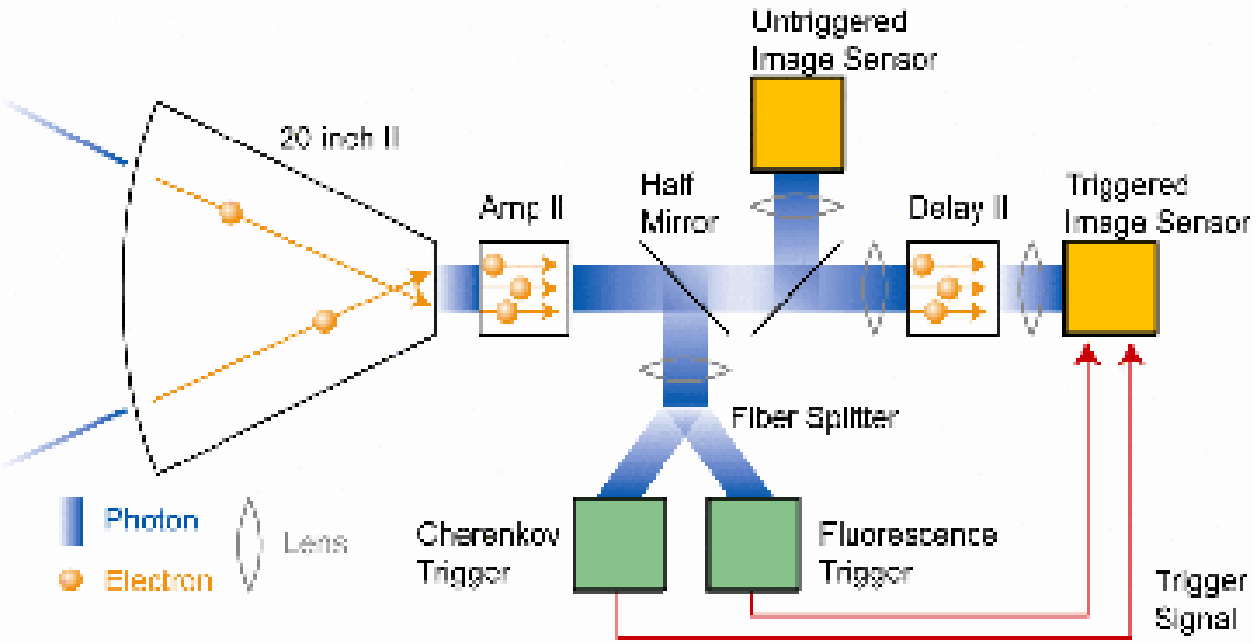}
  \includegraphics[width=0.4\textwidth]{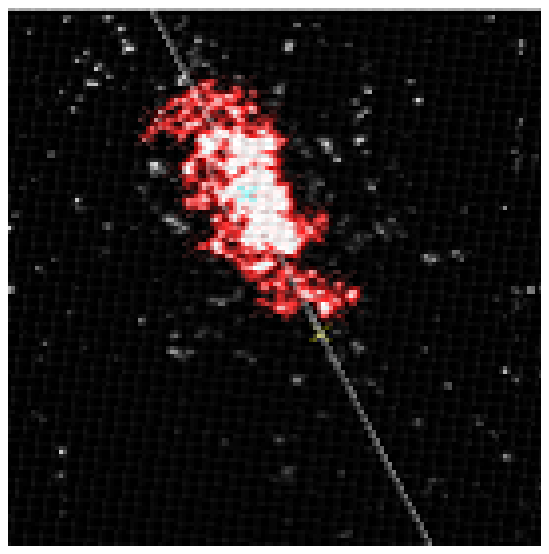}
  \caption{{\bf Top:} Modified Baker-Nunn optic of an ASHRA telescope. {\bf Middle:} Detector and data flow for the various
    types of signal (optical, Cherenkov, fluorescence).
    {\bf Bottom} Example Cherenkov image. All images from \cite{Sasaki:1}.}
  \label{fig:ashra}       
\end{figure}

Another project with large FoV is ASHRA (All-sky Survey High Resolution Air-shower detector), summarised by \cite{Sasaki:1}.
ASHRA has a multi-function focus, aiming to cover the fields of optical/UV photometry, TeV $\gamma$-ray and multi-TeV neutrino
detection, and EeV CR detection. The ASHRA telescopes, under construction 
on Mauna Loa, Hawaii (3300~m a.s.l.) since 2006, employ ultra wide field modified Baker-Nunn optics which provide arcmin focusing 
over a $\sim 40^\circ$ FoV (see Fig.~\ref{fig:ashra} for a photograph). Detected light is amplified by a 20~inch image intensifier
and then diverted to a gated CMOS sensor with 2048$\times$2048 pixels of size 1.2~arcmin (Fig.~\ref{fig:ashra}).
Triggers and gates of different time resolution are used for optical, Cherenkov (TeV $\gamma$-ray and multi-TeV neutrino) 
and fluorescence (EeV CRs) detection. An example Cherenkov image is presented in Fig.~\ref{fig:ashra}. Several ASHRA 
telescopes will be
operated in a local group to provide a full survey of the sky, and eventually, several groups are envisaged separated by about 30~km
on Mauna Loa, Mauna Kea, and Hualali. Its expected integral flux sensitivity (5$\sigma$ in 500~h) is $\sim 10^{-12}$ and $\sim 10^{-14}$
ph~cm$^{-2}$~s$^{-1}$ for energies E$>$1 and 100~TeV respectively. Arcmin angular resolution is also expected in the fluorescence
detection mode.

\subsubsection{Water Cherenkov/Ground Arrays}
There was no contribution at this conference devoted specifically to the technical aspects of the MILAGRO
water Cherenkov detector. However, it is worth highlighting here the fact that
with its discovery of degree-scale multi-TeV $\gamma$-ray emission from the Cygnus and other northern Galactic Plane regions 
(see \cite{Abdo:1,Huentemeyer:1,Walker:1}), MILAGRO has clearly
demonstrated the viability of densely sampling the extensive air shower (EAS) particles, in this case, via their Cherenkov emission
in water. This has provided a solid launching pad for the HAWC proposal, and impetus for upgrades of the Tibet~AS
ground array outlined later in this summary.

A new ground array recently commissioned (mid 2006) at Yangbajing, Tibet (4300~m a.s.l) is the ARGO-YBJ experiment, 
summarised by \cite{Martello:1}. ARGO-YBJ consists of a dense array ($>$10000) of resistive plate chambers (RPCs) covering
an area $\sim 100\times 100$~m. The RPCs provide a signal proportional to the number of incoming charged EAS particles.
Based on timing properties of the triggered RPCs reconstruction of EAS can be performed. A point source angular 
resolution of $\sim 0.5^\circ$ is indicated based on Moon-shadow observations. To-date a 5$\sigma$ excess has been
seen from the Crab in 290~hr after employing a cut based on the density of triggered detectors (so-called Pads).

\subsection{Spaced-Based $\gamma$-Ray Detectors: Current \& Funded}

\begin{figure*}
  \centering
  \hbox{
    \includegraphics[width=0.51\textwidth]{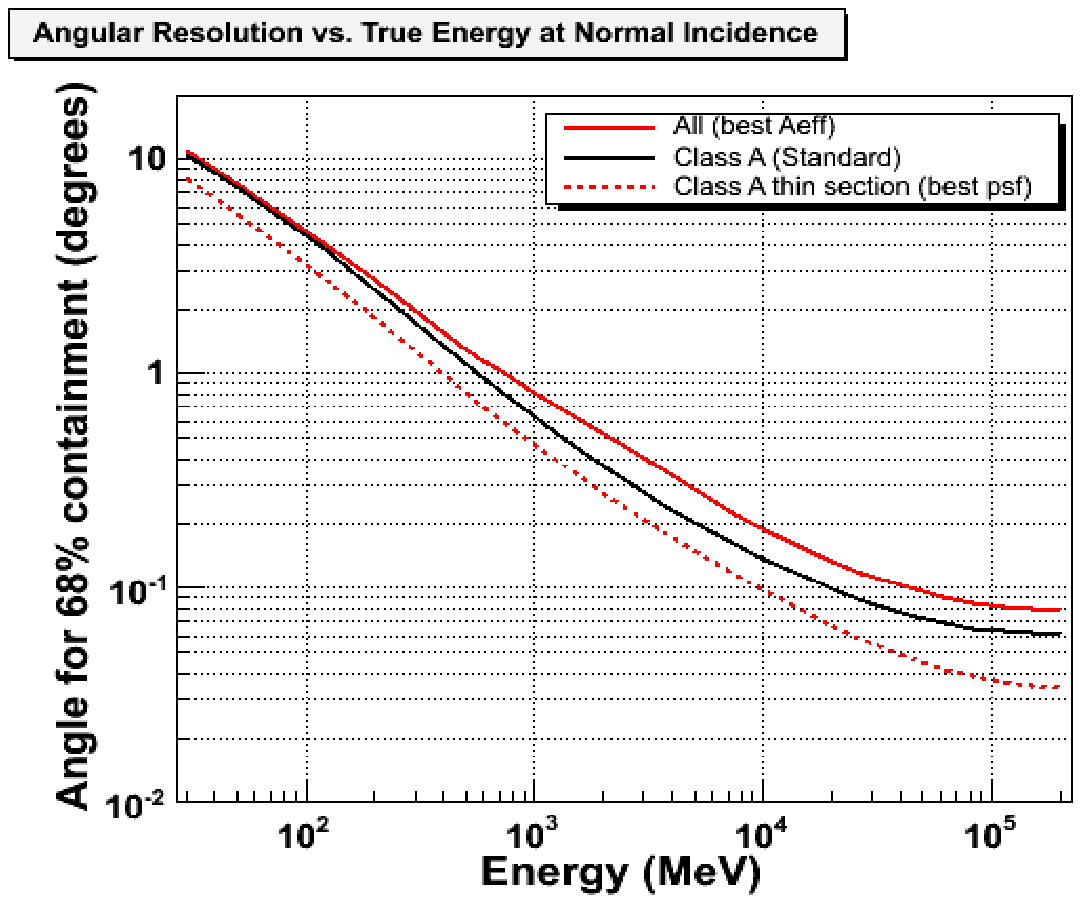}
    \includegraphics[width=0.49\textwidth]{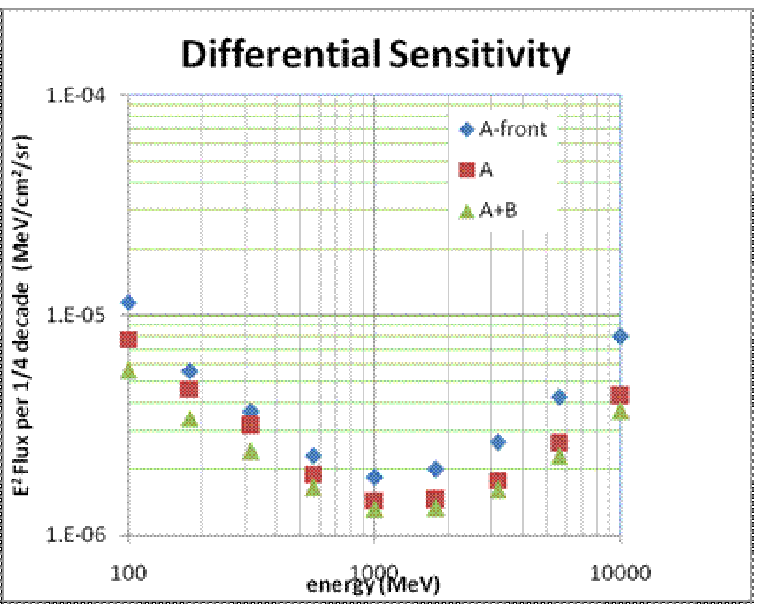}
  }
  \caption{{\bf Left:} GLAST LAT angular resolution vs. energy. {\bf Right} GLAST LAT flux sensitivity for one year 
    observation. Figures are from \cite{McEnery:1}.}
  \label{fig:glast}       
\end{figure*}

In this category we find the eagerly awaited GLAST mission, which will yield a wealth of new astrophysics in the $\sim$20~MeV
to 10's of GeV range, and effectively close the energy gap between space and ground-based
instruments. An overview of GLAST, its timeline, and organisational aspects of data handling 
were presented by \cite{McEnery:1}. GLAST comprises two instruments, the LAT (Large Area Telescope: see \cite{Cohen:1} for details) 
covering the 20~MeV to $>$100~GeV range with a 2.2~sr FoV, and the GBM (GLAST Burst Monitor) operating in the 8~keV to 20~MeV range
with a 9.5~sr FoV. The GBM will trigger on $\sim$215 GRBs per year, with $\sim$70 of these within the LAT FoV. 
The LAT comprises a 4$\times$4 array of dense Si-strip trackers (converting incoming $\gamma$-rays to pairs),
surrounded by an anti-coincidence shield for the rejection of CR particles. At the base of each Si tracker is a CsI calorimeter
for $\gamma$-ray energy estimation and also to aid in CR rejection. The angular resolution of LAT improves strongly with energy
to $\leq0.2^\circ$ for energies above 10~GeV, and its flux sensitivity is a factor 50 or more better
than that of EGRET (Fig.~\ref{fig:glast}). As a result of LAT sky surveys the number of GeV sources is expected to increase
by a factor 10 with sub-arcmin localisation of bright sources.     
At the time of the conference both instruments had completed their final lab and environment testing and had
been integrated with the spacecraft. Transport to the launch site and integration with the Delta~II rocket launch vehicle 
is expected in the latter half of this year. The anticipated launch is in early 2008.
After a $\sim$60~day checkout immediately after launch, the first year sky survey will commence. Within this timeframe, re-positioning
to cover bright bursts will be carried out, and LAT raw data will remain proprietary to the GLAST collaboration. However all GBM data 
and high level information (flux, spectra, location) for LAT bursts and selected sources monitored by LAT will be made public
(this monitoring list is available at http://glast.gsfc.nasa.gov/ssc/data/policy/ LAT\_Monitored\_Sources.html). After this first year,
GLAST observations will be driven by peer-reviewed proposals from guest investigators (GI) with a default state being a sky survey mode.
GLAST Multiwavelength coordination policy was outlined by \cite{Carson:1} and details of the GLAST GI programme
can be found at http://glast.gsfc.nasa.gov/ssc/proposals/. The first phase deadline has already passed at the time of writing
however further GI phases will be announced. In addition to this, strong links with IACT arrays are under development in order to
maximise the astrophysics from transient $\gamma$-ray sources. 

\subsection{Ground-Based $\gamma$-Ray Detectors: New Ideas \& Proposals}

The success of H.E.S.S. in increasing the number of TeV $\gamma$-ray sources (eg. see the
interactive web-based TeV catalogue {\em TeVCaT} summarised by \cite{Wakeley:1} at http://tevcat.uchicago.edu) has 
prompted many in the fields of particle astrophysics and high energy astrophysics to seriously consider the next step in 
ground-based $\gamma$-ray astronomy. A major component of the OG.2.7 sessions were devoted to future
$\gamma$-ray IACT arrays. 

\begin{figure*}
  \centering
  \hbox{
    \includegraphics[width=0.51\textwidth]{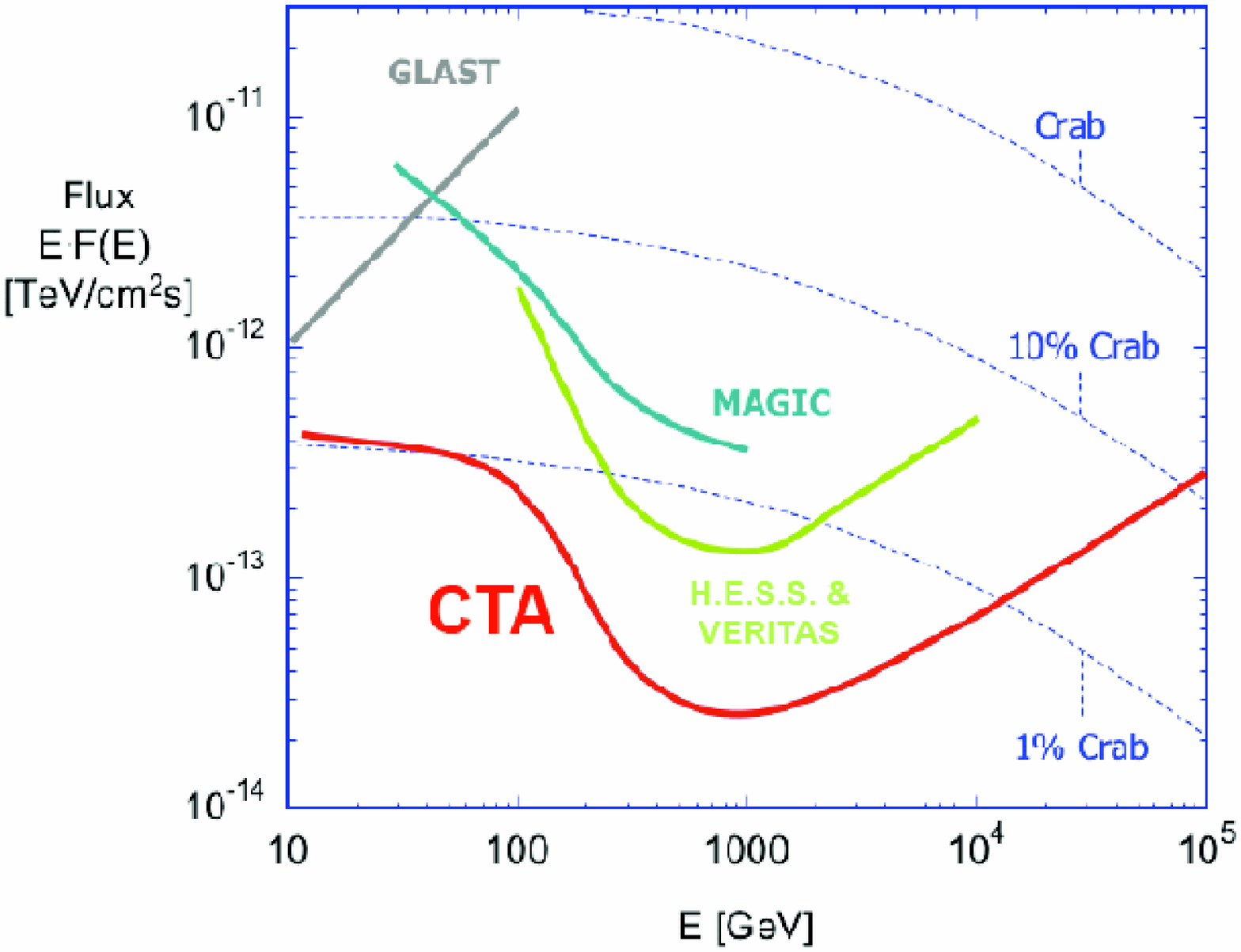}
    \includegraphics[width=0.49\textwidth]{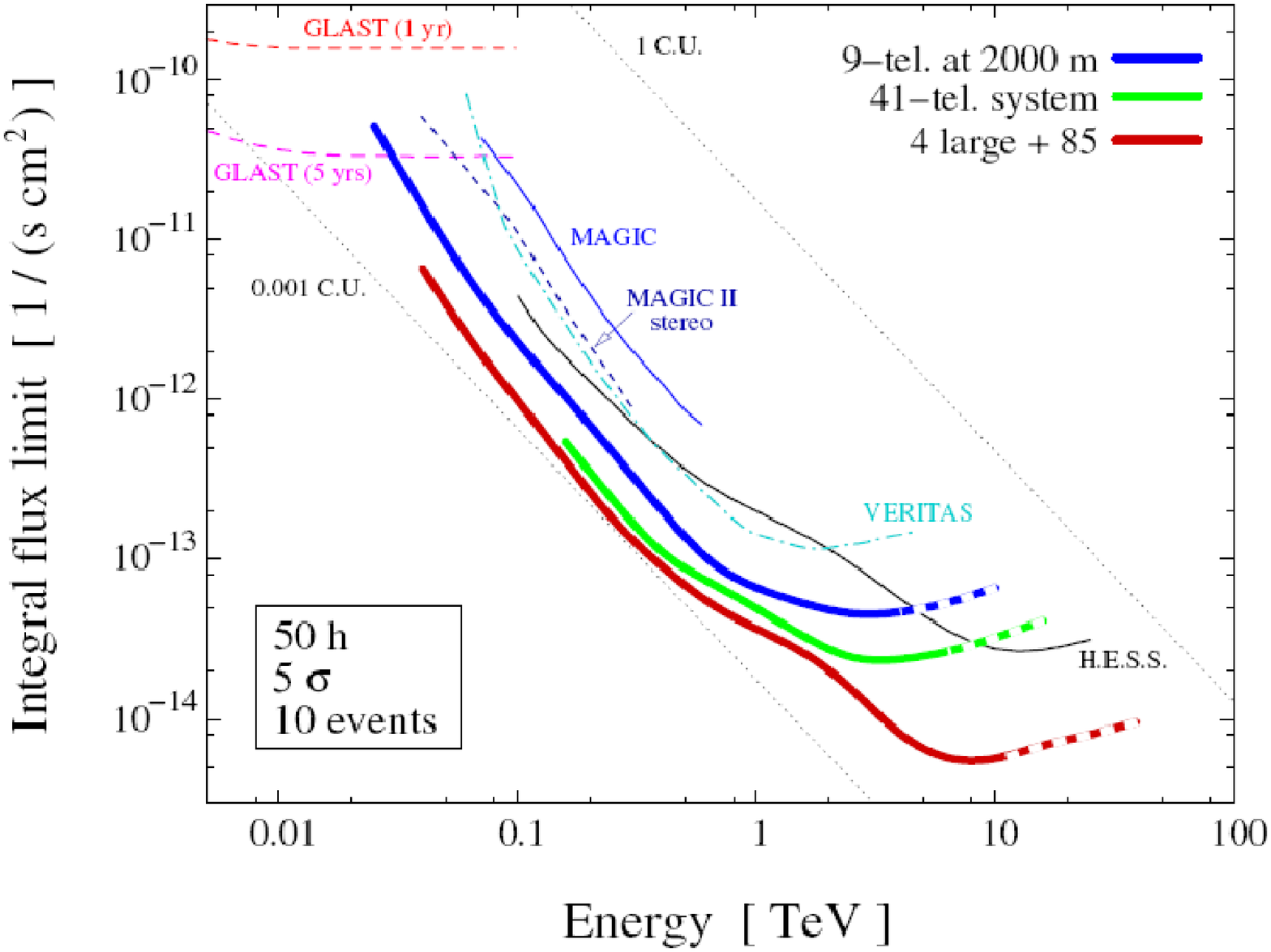}
  }
  \caption{{\bf Left:} Desired flux sensitivity of CTA. {\bf Right} Integral flux sensitivities for several telescope combinations
    considered in CTA simulations so far -- 9-tel (group 1), 41-tel (group 2), 85+4-tel (group 3). See text for explanations of telescope
    groups. Figures are from \cite{Hermann:1,Bernlohr:1}.}
  \label{fig:cta}       
\end{figure*}

Summarised by \cite{Hermann:1}, the European-led Cherenkov Telescope Array (CTA) project \cite{CTA} is looking at the development 
of new $\gamma$-ray IACT arrays, aiming to realise a factor 10 or more improvement in flux sensitivity 
(to the 1 mCrab level - see Fig~\ref{fig:cta}) 
in the 10's of GeV to $\sim$100~TeV energy range with angular resolution down to the arcmin level in specific energy ranges.
CTA represents a large step beyond H.E.S.S. and MAGIC-II, and over 30 institutes have expressed interest so far.
The science case for CTA was outlined by \cite{Drury:1}.
Specific working groups devoted to science drivers and technical aspects have been setup. 
The philosophy is to make use of existing and proven technologies/methods, such as conventional PMTs (peak QE$\sim$25\%) 
and electronics, and to
have the telescopes operate as an open public observatory available to the scientific community. Northern and southern
sites are presently foreseen with a Galactic source emphasis in the south, and extragalactic emphasis in the north, although
there will naturally be considerable overlap. 
MC studies are underway \cite{Bernlohr:1}, and are considering various arrangements of telescopes (at $\sim$2000~m a.s.l.) 
and telescope size combinations, concentrating first on the low (10~GeV) to medium (1~TeV) range. 
Preliminary flux sensitivity results for telescope groups (1) 9$\times$420~m$^2$, (2) 41$\times$100~m$^2$ and (3)  
85$\times$100~m$^2$+4$\times$600~m$^2$ were reported (Fig.~\ref{fig:cta}). Group (3), with the largest number of telescopes,
spread over a 1~km$\times$1~km area is a factor $\sim$10 more sensitive than H.E.S.S. Further work is ongoing to optimise
CR background rejection and stereo reconstruction, and performance improvements beyond these results are anticipated.
Cost estimates for CTA so far place the southern facility at $\sim$100~MEuro, and the northern facility at $\sim$50~MEuro. 
Emphasising its importance to future European science, CTA was shortlisted in the ESFRI (European Strategy Forum on Research 
Infrastructures) 2006 reports, and a design study proposal was submitted to the European FP7 funding round.
 
The US-based White Paper \cite{Krawczynski:1,WhitePaper} sets out to map the path to a future ground-based
$\gamma$-ray instrument. It is the result of several meetings in the US devoted to the future of ground-based $\gamma$-ray
astronomy in that country (Malibu 2005, Santa Fe 2006, Chicago 2007). Initiatives resulting from these have been the White
Paper itself (due in late 2007) and R\&D proposals for new telescopes. Arising from this the AGIS concept  
(Advanced Gamma-Ray Imaging System) has gained momentum. Several working groups are focused on 
specific aspects of the Paper such as different source types
and technical aspects. The White Paper is overseen by an editorial board, comprising eight prominent members of the community.
The next meeting {\em Towards the Future of Very High Energy Gamma-Ray Astronomy} 
(http://www-conf.slac.stanford.edu/vhegra/) taking place at SLAC in November 2007, will further advance plans for AGIS and White
Paper content.

\begin{figure}[p]
  \centering
  \includegraphics[width=0.45\textwidth]{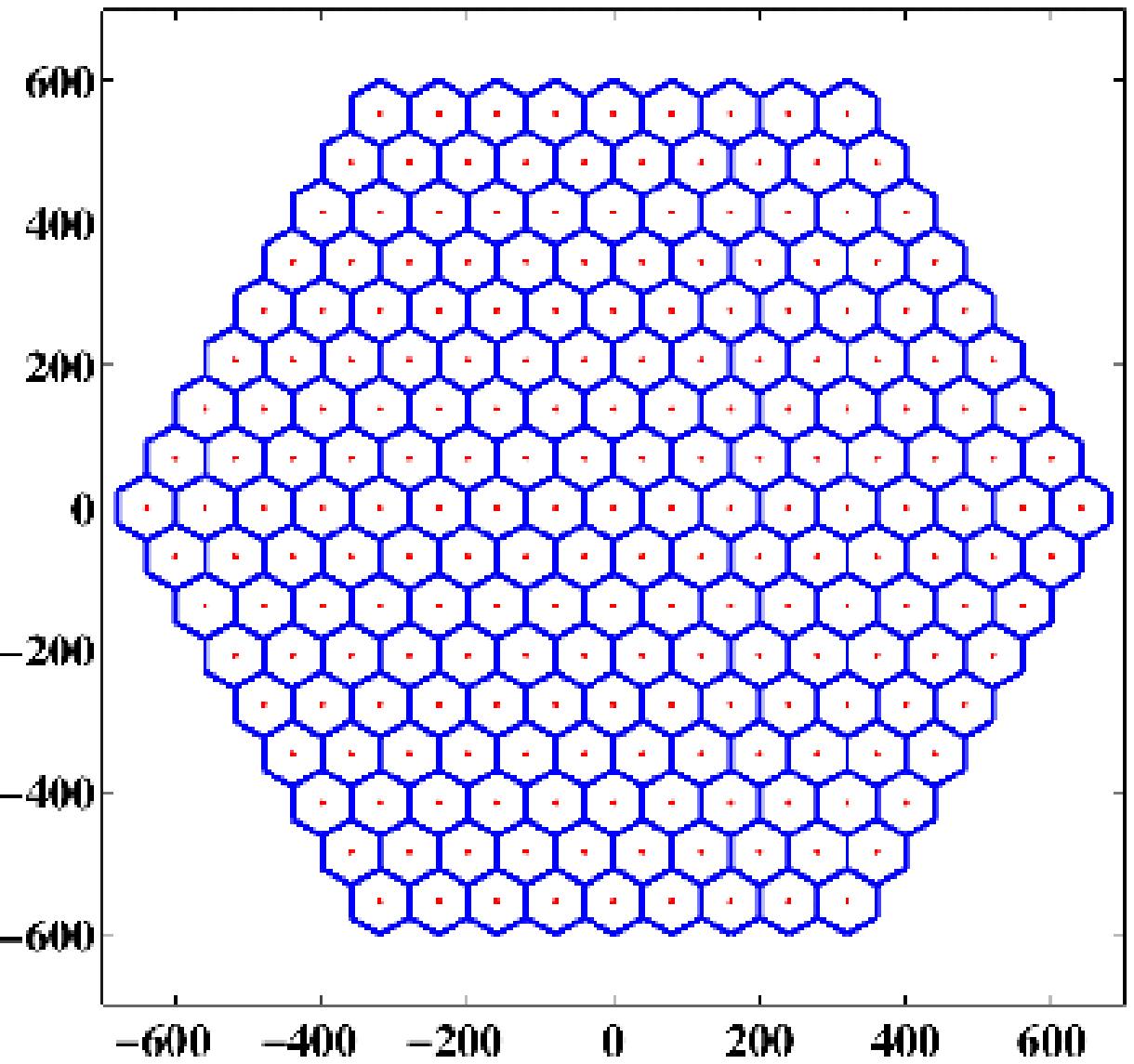}
  \caption{Array layout for 217$\times 10$~m diameter telescopes with 80~m spacing \cite{Fegan:1}.}
  \label{fig:farray}       
\end{figure}

\begin{figure*}[p]
  \centering
  \hbox{
    \includegraphics[width=0.51\textwidth]{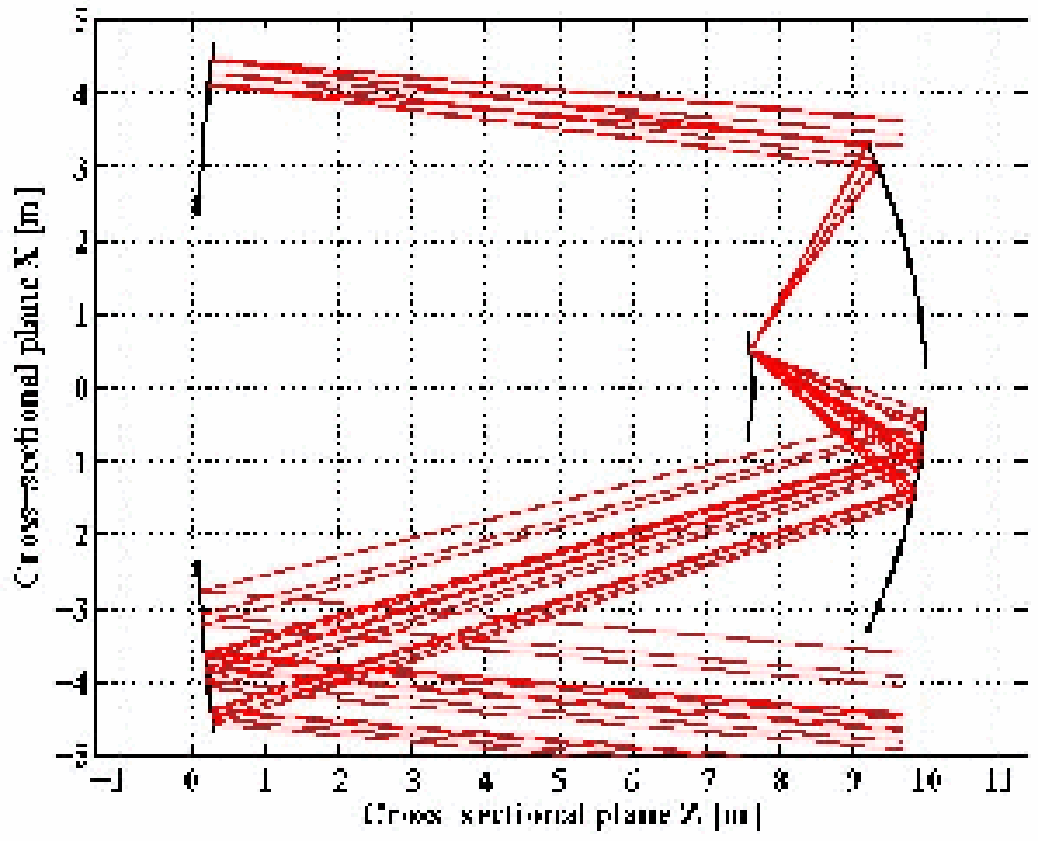}
    \includegraphics[width=0.5\textwidth]{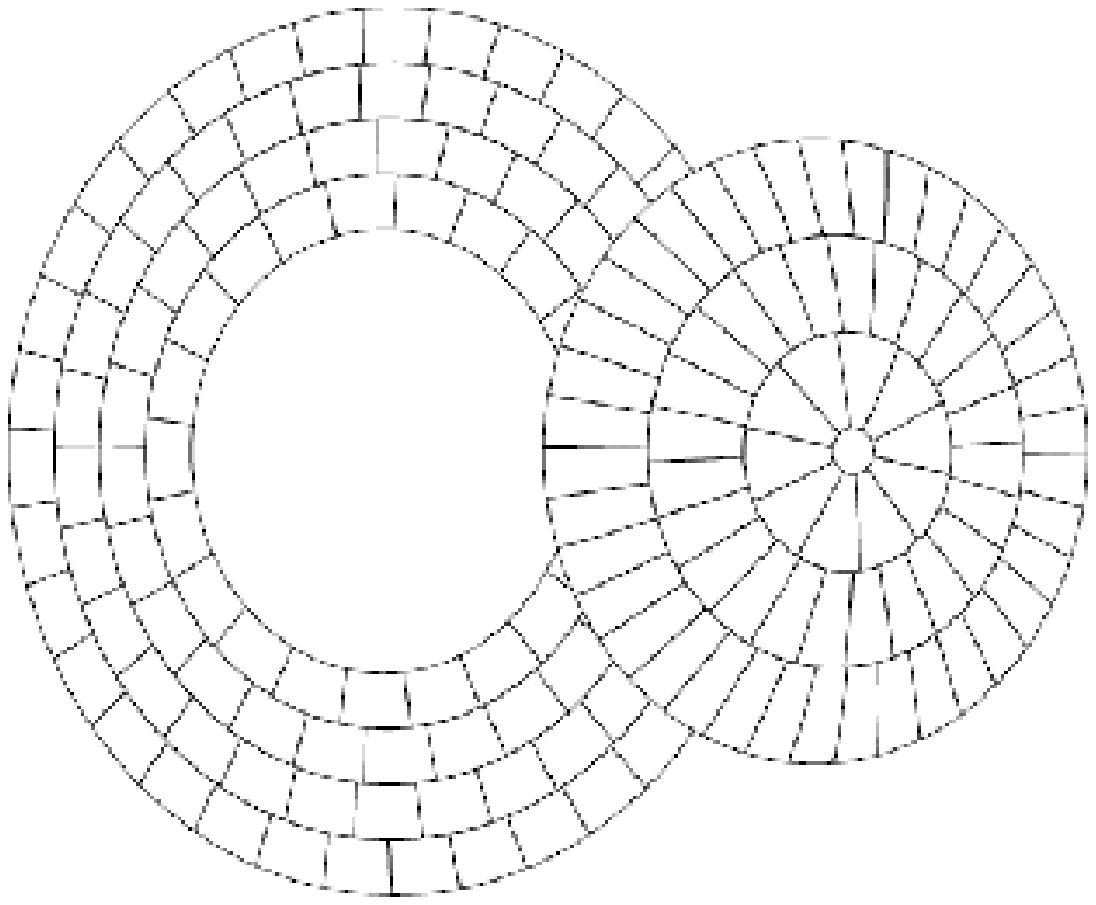}
  }
  \includegraphics[width=1.0\textwidth]{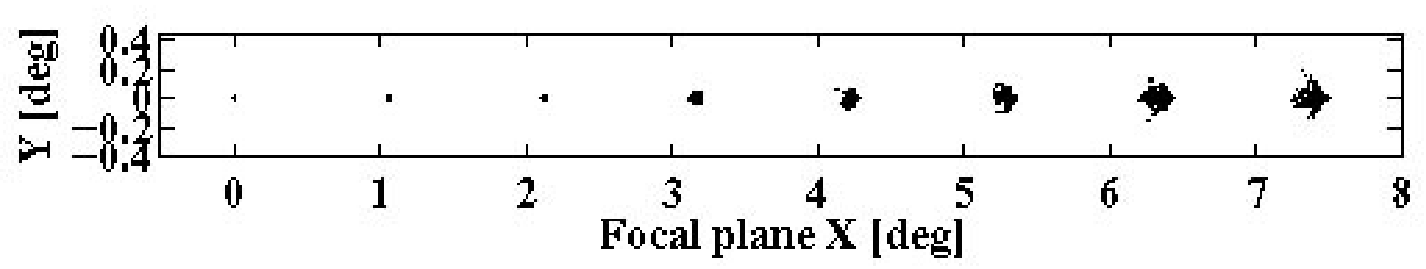}
  \caption{{\bf Upper left:} Off axis ray tracing for the  Schwarzschild-Couder (SC) outlined by \cite{Fegan:1,Vassiliev:1}.
  {\bf Upper right:} Segmented primary and secondary mirror layouts. {\bf Bottom} Off-axis focusing of a particular SC design.}
  \label{fig:scoptics}       
\end{figure*}

Working towards AGIS, \cite{Fegan:1} outlined results from MC simulations of a dense telescope array covering a 1~km$^2$ on the
ground. 217$\times 10$~m diameter telescopes were arranged on a hexagonal grid with 80~m spacing (Fig.~\ref{fig:farray}). 
This small telescope
spacing ensures that each event is well-sampled -- on average three telescopes see the same event, maintaining high performance
(in terms of angular resolution and CR background rejection) across its desired energy range of $<$50~GeV to ~$\sim$1~TeV. 
Their simulations also showed that optimal sensitivity was obtained using camera pixel sizes of 2 to 4~arcmin, for the 
energies 40 and 100~GeV tested. This small pixel size prompted investigation into new type of mirror optics beyond the 
conventional prime-focus types currently in use. Optics based on the Schwarzschild-Couder (SC) design employing a secondary
reflecting surface and curved focal plane have been studied in detail \cite{Vassiliev:1}.
The segmented primary and secondary mirror are aspheric and are formed to correct for spherical and coma aberrations. 
Ray tracing was used to find mirror parameters that minimised astigmatism and effective area losses vs. off-axis angle.
Fig.~\ref{fig:scoptics} depicts an off-axis ray tracing situation, mirror segmented patterns, and the arcmin 
focusing performance out to 7$^\circ$ off-axis. An additional consequence is that the focal plane scale is much reduced (by a factor
$>2$) compared to a prime-focus system, bringing into play multi anode PMTs with small pixel pitches 
(and considerably less per pixel cost compared to single PMTs). 
than single PMTs).  

An additional optics design based on Schmidt optics with a Fresnel corrector was summarised by \cite{Mirzoyan:1}.
Practical aspects of construction of the Fresnel lens, and primary mirror tessellation were also discussed.
Focusing of $\sim$1~arcmin was demonstrated in their f/0.8 system out to off-axis angles of $\sim$7.5$^\circ$.
This, and the SC design discussed earlier, offer promising ways to realise ultra-wide FoV IACTs for the next step
in $\gamma$-ray astronomy. 
  
\begin{figure}
  \centering
  \includegraphics[width=0.48\textwidth]{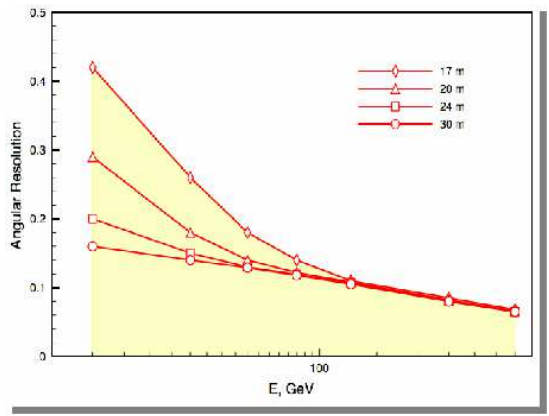}
  \caption{Angular resolution for a three telescope system vs. telescope diameter \cite{Konopelko:1} 
    (telescope separation is 50~m).}
  \label{fig:konopelko}       
\end{figure}

Continuing with the future IACT design, \cite{Konopelko:1} has looked at the design of a three telecope array with an emphasis
on sub-100~GeV energies. Telescope diameters of 17 to 28~m (with 3$^\circ$ FoV camera with 0.07$^\circ$ pixel sizes) with
separation 50 to 100~m
were considered giving peak detection $\gamma$-ray rates at
50 and 25~GeV respectively (for 50 and 80~m separation). CR background rejection based on scaled width and length, as well as 
angular resolution 
were studied for the various telescope sizes. Angular resolution was found to improve with telescope size, varying primarily
in the $E<$50~TeV regime (Fig.~\ref{fig:konopelko} presents results for a 50~m separation). 

Idealised detectors (perfect optics and cameras) were considered by \cite{Bugaev:1} in their optimistion 
study of a 3$\times$18~m diameter telescope array at a moderately high altitude (2700~m a.s.l.) --- based on a possible site of a 
northern IACT array. Similar to the previously described study, the focus was on low energy $\gamma$-rays (here 
3 to 2000~GeV). They found that a large camera FoVs of $\sim 8^\circ$ provide improved point-source sensitivity and pixel sizes 
up to $\sim$0.15$^\circ$ provide similar performance as with a perfect camera. Further work is in progress to extend the number of
telescopes.

Studies of the stereo reconstruction of $\gamma$-ray Cherenkov images was a focus of work presented by \cite{Sajjad:1}
in their simulations of arrays of up to 100 telescopes. Telescopes of 12.5~m and 30~m diameter were considered at 1800~m and
3000~m a.s.l respectively. Direction reconstruction was based on maximum likelihood method incorporating camera pixels from all
telescope simultaneously. 
Optimal telescope spacings (for optimal angular resolution better than 0.1$^\circ$) over the 10~GeV to 10~TeV regime appear in 
the range 100 to 200~m . The
lower altitude site is beneficial for higher energy events due to the fact that shower development may not have fully completed
at 3000~m towards higher energies.

Of interest is the MACE (Major Atmospheric Cherenkov Experiment) telescope is envisaged for the 4200~m a.s.l. 
site in Hanle \cite{Acharya:1}. At present the wavefront sampling experiment HAGAR is currently operating at this
site. MACE is a single large 21~m diamter telescope of very similar design to the MAGIC telescope. A 5$^\circ$ FoV
camera with 0.1$^\circ$ and 0.2$^\circ$ pixels is being considered. Given the very high altitude of this site, the MACE
telescope could operate with a threshold approaching 20~GeV. Funding permitting, construction could get underway by 2010,
with a possible extension to stereoscopic operations a couple of years later.

\begin{figure*}
  \centering
  \hbox{
    \includegraphics[width=0.28\textwidth]{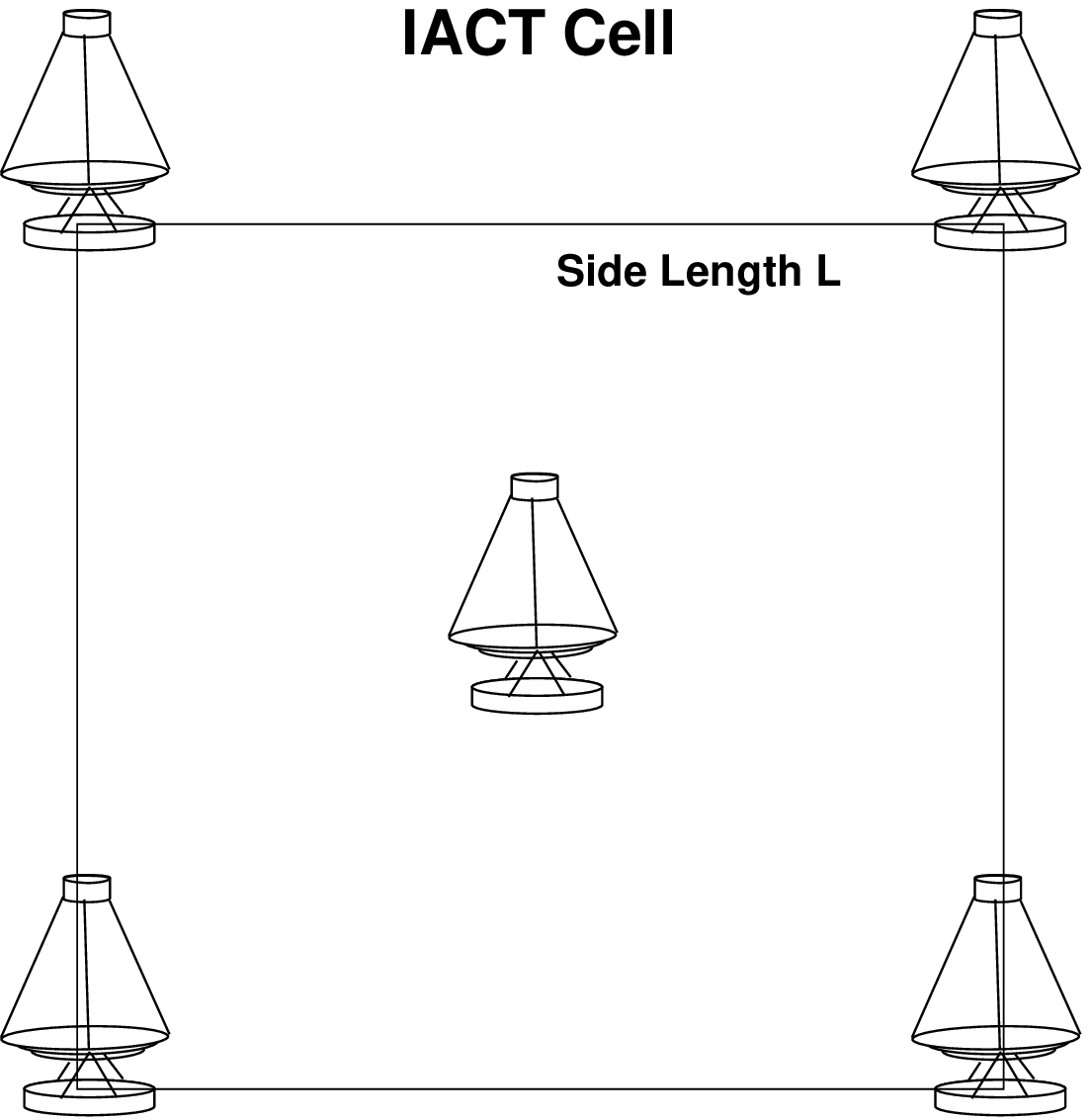}
    \includegraphics[width=0.70\textwidth]{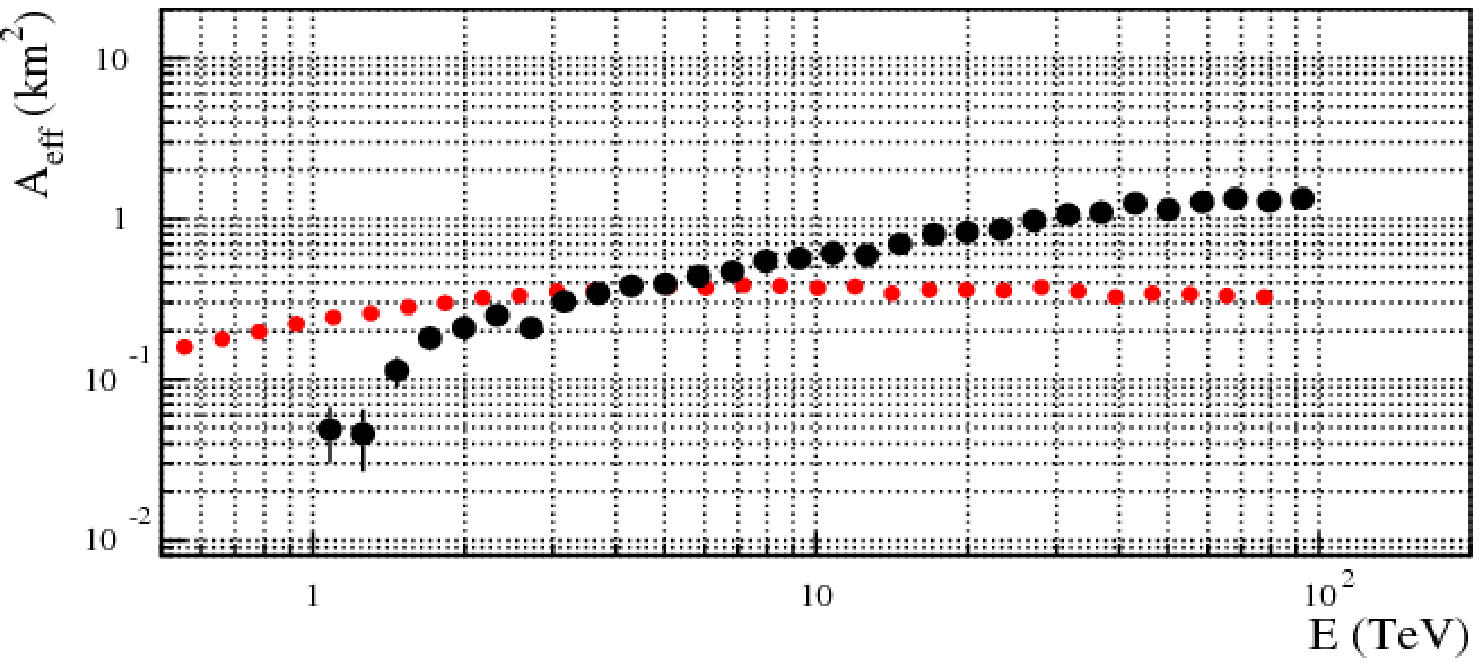}
  }
  \caption{{\bf Left:}
    Layout of an IACT cell of 5 telescopes as simulated for {\em TenTen}. Results for $L=300$~m were presented.
    {\bf Right:} Effective area A$_{\rm eff}$ (km$^2$) vs energy $E$ (TeV) for the 5-telescope cell of side length 
    $L$=300~m, at 200~m a.s.l. (large black solid circles). The effective area achieved 
    by H.E.S.S. is shown as small red solid circles. In both cases, no selection cuts on $\gamma$-ray events are applied.}
  \label{fig:tenten}       
\end{figure*}

With an emphasis mainly on higher energies ($>$10~TeV), \cite{Rowell:1,Stamatescu:1} outlined the {\em TenTen} IACT array
concept. An array of modest-size telescopes (4 to 6~m diameter) coupled to large FoV cameras (up to 10$^\circ$ as permitted
by conventional optics such as the Davies-Cotton design) with large telescope spacing ($>$200~m) is the basis behind {\em TenTen}.
The name is derived from the $>$10~km$^2$ effective collection area required in the $>$10~TeV range for sufficient sensitivity
to discover and study multi-TeV sources approaching the mCrab level. 
MC simulations of a cell of 5x23~m$^2$ telescopes with separation 300~m on a side was presented. Cameras of 8$^\circ$ FoV with
1024$\times 0.25^\circ$ pixels were employed. The cell philosophy is based on earlier experience with the 5-telescope
HEGRA IACT-System and later H.E.S.S. The simulations suggest an effective area exceeding 1~km$^2$ for energies $>$30~TeV 
can be achieved (Fig.~\ref{fig:tenten} presents the layout and effective area curve), as well as similar angular resolution 
and CR background rejection as obtained by HEGRA and H.E.S.S. in their respective energy regimes. The large effective area 
results mainly from the camera FoV, which permits $\gamma$-ray events to trigger out to core distances $>$600~m. 
The relative wide telescope spacing appears sufficent to provide stereosopic views of multi-Tev $\gamma$-ray events. 
Each cell can operate independently if spaced sufficiently apart (eg. $>$1~km) such that the multi-cell performance can be 
easily extrapolated. A ten cell system would then provide the necessary 10~km$^2$ collection area, and based on its collection
area improvement over H.E.S.S., could operate with a flux sensitivity roughly a factor 5-10 better in the 10 to $>$100~TeV
range. Present simulation place the telescope at a near sea-level altitude (200~m a.s.l.) where a collection area improvement 
can be gained compared to higher altitude sites. Futher optimisation of cell layout, trigger conditions, CR background rejection
and site selection are currently underway.

Design studies for 1 to 100~TeV $\gamma$-ray detection were also outlined by \cite{Colin:1}.

\begin{figure*}
  \centering
  \includegraphics[width=1.0\textwidth]{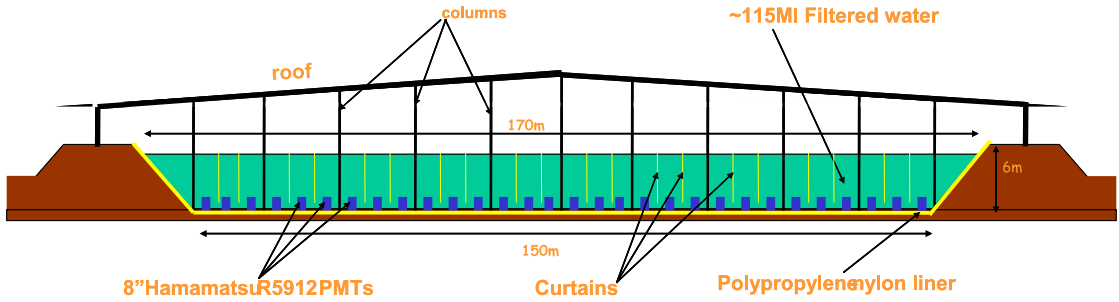}
  \caption{HAWC schematic indicating location of PMTs and isolation curtains \cite{Gonzalez:1}.}
  \label{fig:hawc}       
\end{figure*}

\begin{figure}
  \centering
  \includegraphics[width=0.5\textwidth]{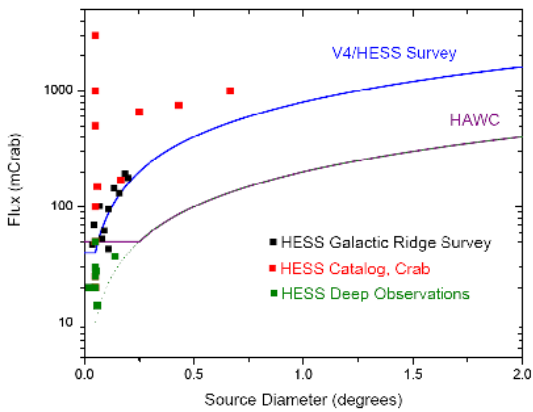}
  \caption{Two year survey sensitivitiy of HAWC in comparison with several H.E.S.S. surveys. \cite{Gonzalez:1}}
  \label{fig:hawc_survey}       
\end{figure}

The HAWC (High Altutide Water Cherenkov Telescope) is the planned water Cherenkov replacement of MILAGRO. Improving on MILAGRO's design,
the key new aspects of HAWC (summarised by \cite{Gonzalez:1}), are its high altitude 
(4100~m at the Sierra Negra site in Mexico \cite{Carraminana:1}) to improve access to shower
particles, and optical isolation of the PMTs in deep (6~m) water so that only one PMT layer is sufficient 
for CR background rejection (Fig.~\ref{fig:hawc}). The optical isolation also reduces accidental triggers. The HAWC sensitivity 
is a factor
$\sim$15 better than MILAGRO, obtaining a 5$\sigma$ detection on the Crab in one day, and a 30~mCrab survey of the northern
sky in 2~yr (Fig.~\ref{fig:hawc_survey}) with an angular resolution in the 0.25$^\circ$ to 0.4$^\circ$ range.  
Given its all sky FoV and 24~hr duty cycle, HAWC would be an ideal complement to future IACT arrays and will no doubt lead
to new discoveries of large-scale TeV sources.
A key announcement at this conference was that Sierra Negra  was officially confirmed as the site for HAWC.

\begin{figure}
  \centering
  \includegraphics[width=0.48\textwidth]{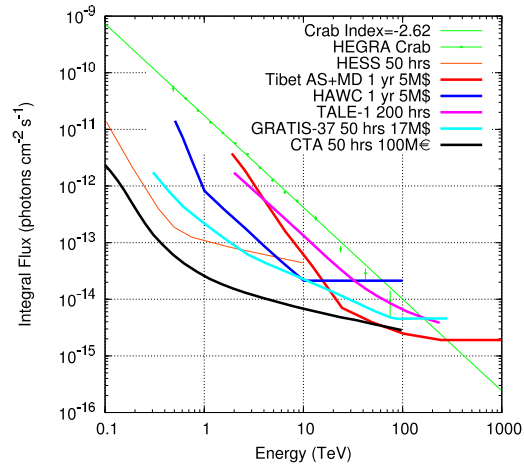}
  \caption{Integral flux sensitivity of the proposed Tibet AS + muon detector (MD) array after 1~yr observation (red line).}
  \label{fig:tibetII}       
\end{figure}

At the same site as ARGO-YBJ, the Tibet Air Shower array has been in operation since 1990 and has undergone several upgrades.
Presently the array consists of 789 plastic scintillator detectors of area 0.5$^2$~m each on a 7.5~m grid spacing. The total
array area is 37000~m$^\circ$. Current methods to discriminate between $\gamma$ and CR events in the detection of the Tail-In, 
Cygnus, and Crab sources were outlined by \cite{Amenomori:1}.
Motivated by the H.E.S.S. Galactic plane results, and to improve the CR background rejection beyond the current modest levels, 
it is planned to install a 
$\sim 200 \times 50$~m$^2$ water Cherenkov muon detectors around the scintillator array \cite{Amenomori:2}. 
Monte Carlo simulations indicate that for energies above 10~TeV, a cut on the level of Cherenkov signal provides a 
CR survival efficiency of $< 10^{-2}$ in comparision to a $\gamma$-ray efficiency of $>$0.5. The integral flux sensitivity
of less than 10$^{-14}$~ph~cm$^{-2}$~s${-1}$ for $E>100$~TeV (Fig.~\ref{fig:tibetII}) is indicated.

\subsection{Spaced-Based $\gamma$-Ray Detectors: New Ideas \& Proposals}

The two new space-based $\gamma$-ray detectors discussed at this conference centred on polarimetry studies, and a new
$\gamma$-ray imager in the $<$100~MeV regime.

POLAR \cite{Lamanna:1} is designed to measure the polarisation of $\gamma$-rays in the few~keV to few~MeV range.
Its science focus is Gamma-Ray Bursts (GRBs) in which polarisation has been flagged a key discriminat of several
models. The core of the detector is made up of a plastic scintillator array of 240$\times$240 elements. Each element is optically
isolated and 6$\times$6 elements are combined to form a so-called bar. Determining the polarisation angle is based on the
coincident detection of Compton recoil electrons and photons in bars of detectors. Vetos to reject CR triggers are incorporated.
Successful laboratory tests of a single bar have pushed the proposal of full space-borne mission. It is expected that POLAR
will be able to detect polarisation fractions $\geq$10\% in those GRB with total fluence 10$^{-5}$~erg~cm$^{-2}$. Roughly 10
events of this type per year could be detected.  

\begin{figure}
  \centering
  \includegraphics[width=0.48\textwidth]{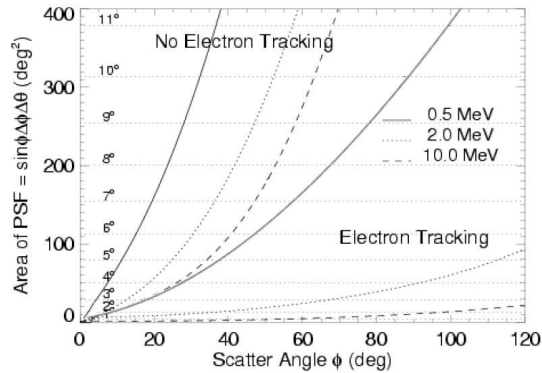}
  \caption{Angular resolution (area and radius) vs. incident photon angle in the 3-DTI with and without electron tracking 
    \cite{Hunter:1}. Results for several photon energies are shown.}
  \label{fig:3dti}       
\end{figure}

The three dimensional track imager (3-DTI) is aimed for 0.3 to 50~MeV $\gamma$-ray astronomy, and will provide an order
of magnitude improvement in sensitivity over COMPTEL/CGRO. Suggested a concept for NASA's Advanced Compton Telescope,
3-DTI was outlined by \cite{Link:1} with the science case presented by \cite{Hunter:1}. 3-DTI is a large volume time projection
chamber (TPC) with 2D gas microwell detector (MWD) readout. The TPC volume is bound by a drift electrode at the top, and my the
MWD at the bottom. The ionisation tracks from Compton-scattered and/or pair producted electrons (resulting as $\gamma$-ray enter
the TPC), drift down to the MWD and are reconstructed in 2D. The time profile of the tracks enables a 3D reconstruction. 
Such reconstruction for photons undergoing Compton scattering is that their arrival directions are reduced to an arc instead 
of a circle, thereby considerably improving the angular resolution (Fig.~\ref{fig:3dti}). Lab tests of a prototype version
have been successfully carried out \cite{Son:1}, paving the way for a balloon and/or space borne version.

\section{\boldmath OG~2.5: Neutrinos -- Gamma-Ray-Related Issues}

The tight coupling between neutrinos and $\gamma$-rays was emphasised in several contributions in this session.
In particular, the well-established $\gamma$-ray fluxes from H.E.S.S., EGRET, and others, permit predictions of the
detection rates in current and proposal neutino detectors. As well as this, several contributions provided status reports
and plans for coordinated $\gamma$-ray/neutrino observations. For status reports see \cite{Gaisser-RAPP}.

\cite{Stegmann:1} summarised neutrino detection rates in KM3NeT (the proposed neutrino detector in the Mediterranean) 
based on H.E.S.S. Galactic source fluxes. A (Sybill-based) parameterisation of secondary pion and other particle 
production was used along with the suggested effective area of KM3NeT. Additional assumptions (not necessarily applicable to
sources where leptonic emission is considered the most viable case) concerning the H.E.S.S.
sources were that no non-hadronic component was present, the radiation density at the source was low, 
and there is a low magnetic field. The best signal to noise (atmospheric $\nu$) ratios for 5~yr observation 
were indicated for the bright SNR RX~J1713$-$3946 (2.6 to 6.7 source event over a background of 8.2) and the nearby plerion 
Vela-X (5 to 15 evnts over a background of 4.6). Many of the other H.E.S.S. source have fluxes a factor 5-10 lower and so
present more difficult tasks. However, the assumption of low in-source radiation density would not necessarily apply to the
compact binaries, leading to their $\gamma$-ray fluxes acting more as lower limits on the potential neutrino fluxes one
could see. A key point is that given the generally hard source and soft atmospheric neutrino spectra, 
and detector effective area behaviour with energy, better signal to noise ratios are expected at higher energies $>5$~TeV or so.
\begin{figure}
  \centering
  \includegraphics[width=0.48\textwidth]{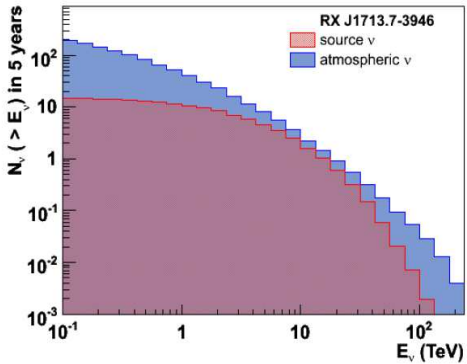}
  \caption{Neutrino event count (in 5~yr) predicted from RX~J1713$-$3946 (source $\nu$) and atmospheric neutrino count vs.
    energy \cite{Stegmann:1}.}
  \label{fig:stegmann}       
\end{figure}

Based on the distribution of known SNRs, and gas in the Galactic plane, CR diffusion properties (propagation in turbulent
magentic fields) \cite{Grasso:1} presented calculations of the diffuse neutrino and $\gamma$-ray flux at different energies. 
Their calculation were able to match EGRET measurements in $<$10~GeV regime, and provide considerably more narrow
latitude distributions in the $>$1~TeV regime. Their 
predicted neutrino flux in the Galactic Centre region for $E>1$~TeV of 4.2$\times 10^{-11}$~cm$^{-2}$~sr$^{-1}$~s$^{-1}$ 
would appear difficult with KM3NeT, and hence require a CR enhancement.

\cite{Bernardini:1} outlined the target of opportunity (ToO) setup between AMANDA-II and MAGIC. 
MAGIC $\gamma$-ray observations are triggered if AMANDA observes neutrino events close to pre-defined sources.  
The list includes AGN and X-ray binaries). Test observations on a few sources was carried out in Sept. to Nov. 2006
in order to assess the feasibility of such a programme. Email alerts are sent from AMANDA-II to MAGIC if any neutrino
event is reconstructed with a few degrees of selected sources. MAGIC observations are carried out if possible within
24~h of the alert. The joint probability of observing $n_\nu$ neutrinos and $n_\gamma$ $\gamma$-rays was assessed
to ensure false alarm rate was not too high.  $n_\gamma$ is directly into $p_\gamma$, the probability to observe a particular
$\gamma$-ray flare within the prescribed time limit. Effort is devoted to defining this term based on long-term monitoring
of sources and presently upper limits are given. During the test run several alerts were given (on Mkn~421, 1ES2344+514, 
1ES1959+615, LSI+61303,GRS1915+105) and no coincident $\gamma$-ray events or flares were seen.

In the opposite direction, using $\gamma$-ray flares to contrain time windows in neutrino detectors, \cite{Goodman:1}
discussed the possibility of HAWC-triggered time windows in the IceCube detector. 

\section{\boldmath OG~2.6: Gravitational Waves}

Gravitational wave detection was not a major theme at this conference and only two contributions were given. 
\cite{Ugolini:1} outlined the fact that CR-induced charge on LIGO (designed for 10~Hz to 3~kHz gravity wave detection)
optics could be a major source of noise (from charge motion and dust attraction) in its sensitive frequency range. 
Methods to mitigate this charge build up were discussed.

\section{Conclusion}

Some key conclusions on these sessions primarily devoted to the detection of $\gamma$-rays and neutrino can be outlined as follows:

\begin{itemize}
\item Ground-based $\gamma$-ray astronomy is a now a mature field employing two established and complementary
  techniques (a) (Stereoscopic) Imaging Atmospheric Cherenkov Imaging; (b) Water Chereknov detection of air shower particles.
  All of the planned and proposed detectors appear to be making use of at least one of these two techniques.
  
\item The next step beyond H.E.S.S., VERITAS, MAGIC-II etc.. is taking shape via continental-wide organisation efforts
  such as CTA (Europe) and the WhitePaper (USA), along with several other proposals. This is necessary in order to gather 
  community-wide support for the required funding scales of order 100 Million dollar/Euros.

\item GLAST is ready to go and its launch date is not far away. GLAST will no doubt provide a fresh and clearer look at the 
  MeV to GeV Universe.

\item Neutrino event rates in forthcoming detectors (IceCube, KM3NeT) 
  appear to be a few per year from the strongest $\gamma$-ray sources. This would give them a chance to realise discovery
  of extraterrestrial neutrino sources. In addition the first efforts in coordinated neutrino/$\gamma$-ray observations have begun.
\end{itemize}

I wish to thank the organisers of the 30th ICRC for giving me the opportunity to summarise these topics. I also thank in 
particular Simon Swordy and Tom Gaisser for presenting my slides.

\scriptsize


\bibliographystyle{plain}
\bibliography{rowell_rapp}

\end{document}